\documentclass[a4paper,10pt]{article}
\usepackage{amssymb, latexsym, amsmath,color,url}
\usepackage{citesort}

\newcounter{mnotecount}[section]

\renewcommand{\themnotecount}{\thesection.\arabic{mnotecount}}
\newcommand{\mnote}[1]%{}
{\protect{\stepcounter{mnotecount}}$^{\mbox{\footnotesize
$%\!\!\!\!\!\!\,
\bullet$\themnotecount}}$ \marginpar{%\color{red}%
\raggedright\tiny\em
$\!\!\!\!\!\!\,\bullet$\themnotecount: #1} }

\begin{document}

\title{A2: Mathematical relativity and other progress in classical gravity theory --
a session report%
\thanks{Preprint UWThPh-2013-33.}}
\author{
Piotr T. Chru\'sciel{}\thanks{Email  {Piotr.Chrusciel@univie.ac.at}, URL {
http://homepage.univie.ac.at/piotr.chrusciel}}\ \ and
Tim-T. Paetz{}\thanks{Email  Tim-Torben.Paetz@univie.ac.at}   %\vspace{0.5em}  \\  University of Vienna
 \vspace{0.5em}\\  \textit{Gravitational Physics, University of Vienna}  \\ \textit{Boltzmanngasse 5, 1090 Vienna, Austria }}
\maketitle

\vspace{-0.2em}

\begin{abstract}
We report on selected oral contributions to the A2 session ``Mathematical relativity and other progress in classical gravity theory''
of ``The 20th  International Conference on General Relativity and Gravitation (GR20)'' in Warsaw.
\end{abstract}

\maketitle

\section{Introduction}
The talks in the A2 workshop were mainly concerned with progress in the mathematical understanding of general relativity. The session included six invited half-hour talks by, alphabetically,
 Helmut Friedrich (construction of scattering initial data for asymptotically flat space-times), Gustav Holzegel (existence of large classes of dynamical black hole solutions), Jonathan Luk (construction of nonlinear impulsive waves), Hans Ringstr\"om (dynamical stability of cosmological models), J\'er\'emie Szeftel (delicate control of the local dynamics of the Einstein equations) and Robert Wald (stability of black holes and strings), describing spectacular progress on each of the topics.

Out of  many excellent abstracts submitted to the session, eighteen have been selected for  fifteen-minutes oral presentations,  with the remaining  moved to the poster session because of  time constraints. In this note we will shortly report on the talks presented during the session. The topics naturally split into evolution problems, black holes studies, and other;
this report is organized accordingly.

\section{Evolution problems}

In his talk, \textbf{Gustav Holzegel} reported on joint work with Mihalis Dafermos and Igor Rodnianski on the  \emph{Construction of Dynamical Vacuum Black Holes}.
The authors
prove existence of a large class of dynamical vacuum black hole spacetimes whose exterior geometry asymptotes in time to a fixed Schwarzschild or Kerr metric. The spacetimes are constructed by solving a backwards scattering problem for the vacuum Einstein equations with characteristic data prescribed on the event horizon and (in the limit) at null-infinity. The solutions are  parameterized by scattering data, and the class admits the full functional degrees of freedom to specify data for the Einstein equations. There is, however, a catch, that the scattering data are assumed to decay exponentially fast, which is not expected for general such solutions. As a result, the solutions converge to stationarity exponentially fast, with their rate of decay being intimately related to the surface gravity of the event horizon. This is the first such family of solutions without imposing an ansatz on the form of the metric (as done in~\cite{ChruscielRT,DafermosHolzegel,BCS}), or assuming symmetries (as, e.g., in~\cite{ChristodoulouAnnals99}).

\bigskip

In his contribution to the session, \textbf{J\'er\'emie Szeftel} outlined the strategy used, together with Sergiu Klainerman and Igor Rodnianski, to establish \emph{The resolution of the bounded $L^2$ curvature conjecture in general relativity}. In their work the authors consider the Einstein vacuum equations. Recall that the general relativistic initial data  consist of a three dimensional   manifold $\Sigma_0$   together with a    Riemannian  metric $g$ and a symmetric $2$-tensor $k$  verifying the constraint equations. It is natural to ask what are the minimal regularity assumptions on the initial data that guarantee local existence of solutions. The bounded $L^2$ curvature conjecture, proposed in~\cite{PDE}, states that the Einstein equations admit   local    Cauchy developments  for initial data sets   $(\Sigma_0, g, k)$ with locally  finite $L^2$ curvature and locally finite $L^2$ norm of the  first covariant derivatives of $k$. The conjecture has been proved recently in~\cite{KRS}, and is  most probably sharp insofar as the minimal     number of derivatives    in $L^2$  is concerned.

The theorem improves upon previously known results, by exploiting in a very delicate way the full nonlinear structure - the so-called null structure - of the Einstein equations. The  assumptions are geometric, in that they make sense without having to refer to a particular coordinate system, and thus preserve the freedom of choice of gauges. The result may  be viewed as a breakdown criterion, and as such improves all previously known breakdown criteria. In particular it identifies the condition that the full spacetime curvature tensor ${\bf R}$ belongs to $ L^2$ as a candidate for controlling singularity formation. Evidence that the criterion is optimal is provided in the proof that the control of the radius of injectivity of light cones, for solutions of  Einstein vacuum equations, requires ${\bf R}\in L^2$.

\bigskip

The talk of \textbf{Jonathan Luk} concerned \emph{Propagation and interaction of impulsive gravitational waves}, as studied jointly with Igor Rodnianski.
Motivated by the explicit examples of plane impulsive gravitational waves of Penrose and Khan-Penrose, the authors considered the propagation and interaction of impulsive gravitational waves \emph{without any symmetry assumptions}~\cite{LR,LR2}. The problem is studied via solving the characteristic initial value problem for the vacuum Einstein equations, where the initial data have a curvature delta singularity supported on a two-surface. It is proved that the singular initial data give rise to a unique local spacetime solution of the vacuum Einstein equations. The resulting spacetime has a curvature delta singularity on the null hypersurface emanating from the initial singularity and the spacetime is smooth away from this hypersurface.

Luk and Rodnianski have  also studied the characteristic initial value problem representing the interaction of two impulsive gravitational waves. In this case, the characteristic initial value problem is solved with data imposed on two hypersurfaces, each of which has a delta singularity in the curvature supported on a 2-surface. In the spacetime, the curvature delta-singularities propagate along 3-dimensional null hypersurfaces intersecting to the future of the data. To the past of the intersection, the spacetime can be thought of as containing two independent, non-interacting impulsive gravitational waves and the intersection represents the first instance of their nonlinear interaction. The authors were able to analyze the region past the  first interaction, and showed that the spacetime  remains smooth away from the continuing propagating individual waves.

\bigskip

\textbf{Hans Ringstr\"{o}m}'s talk focused on \emph{Future stability of models of the universe.}
More precisely, Ringstr\"om considered a class of cosmological solutions of Einstein-Vlasov equations
consistent with observations~\cite{Ringstrom}. Recall that the currently preferred model of the universe is spatially homogeneous
and isotropic, as well as spatially flat,  with  sources consisting of
ordinary matter (such as dust and radiation), dark matter
(often modeled by dust) and dark energy (for which there are many models,
but the talk only concerned the case of a positive cosmological constant). It is clearly of interest to ask what
happens when perturbing the corresponding initial data; do the perturbed
solutions exhibit behaviour to the future similar to the homogeneous models? The usual assumptions, described
above, also lead to very strong restrictions concerning the spatial topology
(even if one relaxes spatial homogeneity to local homogeneity). It is therefore
natural to enquire about the restrictions on the global topology imposed by
the assumption that every observer in the universe sees the universe as being
close (but not necessarily identical) to one of the standard models.
All the issues above have been adressed for models with Vlasov
matter, for which future stability results of the desired type were presented.
The issue of topology was discussed, the conclusion being that it is not
possible to draw conclusions concerning the global shape of the universe.

\bigskip

During the session, \textbf{Helmut Friedrich} discussed \emph{Radiation fields and vacuum solutions near past time-like infinity}.
Indeed, a natural way to study {\it purely radiative,  asymptotically flat space-times},
generated solely by gravitational radiation coming in from past null infinity, is to analyse the asymptotic characteristic initial value problem for the conformal vacuum field equations  where data are prescribed on a cone ${\cal N}_p$ with vertex $p$ similar to the cone $\{x_{\mu}\,x^{\mu} = 0, \,\,x^0 \ge 0\}$ in Minkowski space. The problem is to be arranged so that the prospective vacuum solution admits  a smooth conformal extension in which the point $p$ becomes past time-like infinity $i^-$, and the set $ {\cal N}_p \setminus \{p\}$, swept out by the future directed null geodesics through $p$, represents past null infinity ${\cal J}^-$.
There arise two distinct issues, namely: 1)    constructing initial data  for the field equations from suitably prescribed free data, and 2) proving existence of solutions. Friedrich addressed the first point.
Due to the non-smoothness of ${\cal N}_p$ at its vertex, the algebra of the field equations is complicated and notions of smoothness require special care when
a full data set is to be derived from a freely given
{\it incoming  radiation field} on ${\cal N}_p$. It is shown in~\cite{friedrich:2013} that for
a given, {\it suitably smooth} radiation field on ${\cal N}_p$ there exists on ${\cal N}_p$ a unique set of fields, comprising the radiation field, which satisfy the {\it transport equations} and the
{\it inner constraints} induced on ${\cal N}_p$ by the conformal field equations. These fields are smooth on the three-manifold ${\cal N}_p \setminus \{p\}$
in the standard sense. The key new result in~\cite{friedrich:2013} is the proof that  the resulting initial data  coincide near $p$ at all orders in a Taylor expansion
with fields which are defined and smooth in the standard sense in a neighbourhood of $p$.
These Taylor expansions are used as a starting point for the existence proof given in~\cite{chrusciel:paetz:2013}.

\bigskip

In his contribution,
\textbf{H{\aa}kan Andr\'{e}asson} presented results, obtained with Markus Kunze and Gerhard Rein, concerning \emph{Rotating, stationary, axially symmetric spacetimes with collisionless matter}.
In astrophysics, matter described as a collisionless gas is used to analyze
galaxies or globular clusters
where the stars play the role of the gas particles and collisions among
these are sufficiently rare to be neglected. The particles
interact only by the gravitational field which the ensemble creates
collectively, and the general relativistic description of such an ensemble
is given by the Einstein-Vlasov system, already considered by Ringstr\"om in his talk described above.
The existence of spherically symmetric steady states to this system
has been known for two decades~\cite{Rein:1993ix}. More recently, the existence
of axially symmetric static solutions of the Einstein-Vlasov system was
established in~\cite{AKR1}. Andr\'easson presented
an extension of this result to the case of axially symmetric stationary
solutions which have non-zero total angular momentum~\cite{AKR2}.
This provides a mathematical model for general relativistic  rotating
asymptotically flat non-vacuum spacetimes. The solutions are obtained via
the implicit function theorem by perturbing off a suitable static and
spherically symmetric steady state of the Vlasov-Poisson system.
Due to the non-vanishing of angular momentum, the system of equations
contains a semilinear elliptic equation which is singular on
the axis of rotation, which is handled  by recasting
that equation as one for an axisymmetric unknown on $\mathbb{R}^5$.

\bigskip

In his presentation,
\textbf{Juan A. Valiente Kroon} discussed work with Christian L\"ubbe on \emph{Anti de Sitter-like Einstein-Yang-Mills spacetimes}.
Such spacetimes have been the focus of renewed interest
since the discovery of numerical evidence of turbulent instability in
evolutions of the spherically symmetric Einstein-scalar field system
with reflective boundary conditions by Bizo\'n and Rostworowski~\cite{BizonRostworowski}. The
question of the local existence of vacuum anti de Sitter-like
spacetimes has been previously analysed by Friedrich~\cite{Friedrich:aDS}. It is then natural to study these problems for the Einstein-Yang-Mills equations with negative cosmological constant; compare~\cite{friedrich:JDG} for
$\Lambda \ge 0$. The choice of the Yang-Mills
field as a matter source is motivated by its amenability to a
treatment using conformal equations~\cite{LueVal13a}, and because it
leads to non-trivial dynamics in the spherically symmetric case. The analysis makes use of
a class of conformally privileged curves, which are \emph{not} conformal
geodesics, and shows that there exists a large class of maximally
dissipative boundary conditions making the initial boundary value
problem well-posed. A  numerical treatment
of the spherically symmetric case has been presented
in~\cite{LueVal13b}.

\section{Black holes}

Black holes belong to the most fascinating objects arising in general relativity, and as such remain a favorite topic of research.
There has been key progress in recent years on issues concerning their stability~\cite{Hollands:2012sf,FKR}, and results on this are reported in Section~\ref{ss16XI13.1} below. There has also been renewed interest in extreme black holes following Aretakis' discovery of their instability~\cite{Aretakis4,Aretakis5}, and results on extreme solutions are presented in Section~\ref{ss16XI13.2}.
Some miscellaneous results are collected in Section~\ref{ss16XI13.3}.

\subsection{Stability}
 \label{ss16XI13.1}

In his talk on \emph{Dynamic and thermodynamic stability of black holes and black branes},
\textbf{Robert Wald} described work with Stefan Hollands~\cite{Hollands:2012sf} on a new criterion for the dynamical stability of black holes in $D \geq 4$ spacetime dimensions with respect to axisymmetric perturbations. Here it should be kept in mind that there have been thousands of papers written concerning the ``thermodynamics of black holes",  based on loose classical analogies and the semi-classical Hawking radiation, with no clear mathematical consequences concerning the behaviour of black holes. The remarkable work by Hollands and Wald shows that ``thermodynamical arguments" can be used to obtain mathematically rigorous information concerning the dynamical behavior of linearized perturbations of black-hole backgrounds, as well as to gain insight into nonlinear dynamical stability of the solutions.
More precisely, the authors proved that linearized dynamic stability is equivalent to the positivity of the canonical energy, $\mathcal E$, on a subspace of linearized solutions that have vanishing linearized ADM mass, momentum, and angular momentum at infinity and satisfy certain gauge conditions at the horizon. They showed that $\mathcal E$ is related to the second order variations of mass, angular momentum, and horizon area by
$$
 \mathcal E = \delta^2 M - \sum_i \Omega_i \delta^2 J_i - \frac\kappa{8\pi} \delta^2 A
  \;,
$$
%,
thereby establishing a close connection between dynamic stability and thermodynamic stability. In fact, thermodynamic instability of a family of black holes need not imply dynamic instability because the perturbations towards other members of the family will not, in general, have vanishing linearized ADM mass and/or angular momentum. However, Hollands and Wald prove that a class of black branes corresponding to thermodynamically unstable black holes are dynamically unstable, as conjectured by Gubser and Mitra~\cite{GubserMitra} (in this context, see~\cite{GubserMitraCounterexample}). They also prove that positivity of $\mathcal E$ is equivalent to the satisfaction of a ``local Penrose inequality''~\cite{FKR}, thus showing that the  local Penrose inequality is necessary and sufficient for dynamical stability.

\bigskip

\textbf{Kartik Prabhu}, speaking on \emph{Gauge, energy and stability of black holes}, presented progress in developing further applications of the general result of Hollands and Wald~\cite{Hollands:2012sf}. His analysis is carried out in the ADM framework with a static background spacetime satisfying the vacuum Einstein equations.
For static spacetimes, the energy separates into a kinetic and potential energy part. The kinetic energy can be shown to be positive using a kind of transverse-traceless decomposition. The potential energy contribution can be simplified using a natural gauge condition. The potential energy is shown to be positive for Minkowski spacetime and for axial perturbations on an axisymmetric spacetime.
The case of Rindler spacetime turns out to be  complicated by boundary terms at the horizon, and showing positivity of potential energy has proven difficult. It is hoped that a resolution for the Rindler case will provide clues how to handle the case of static, and ultimately stationary, black hole spacetimes.

\bigskip

In his recent work with Elisabeth Winstanley,
\textbf{Brien Nolan} considered the question of \emph{Existence and stability of dyons and dyonic black holes in Einstein-Yang-Mills theory}.
Initial work on black hole solutions of the Einstein-Yang-Mills (EYM) equations focussed on spherically symmetric gravity coupled to a purely magnetic $\mathfrak{su}(2)$ gauge field in the asymptotically flat setting. In~\cite{Nolan-Winstanley:2012}, the authors consider the issue of the existence of \textit{dyonic} black hole and soliton solutions of the EYM equations with an $\mathfrak{su}(2)$ gauge group in asymptotically anti-de Sitter spacetime. For dyonic solutions, both the electric and magnetic part of the gauge field are non-zero. These are of interest due to the no-go result of Ershov and Galt'sov, which rules out the existence of such solutions in the asymptotically flat setting. The principal result of~\cite{Nolan-Winstanley:2012} is the proof of the existence of a four (respectively, five) parameter family of static, spherically symmetric black hole (respectively, soliton) solutions of the EYM equations with a negative cosmological constant. The proof relies on (i) the construction of local solutions at the horizon (respectively, regular centre); (ii) establishing the manifold structure of the family of local solutions and (iii) the application of the Banach space implicit function theorem to `globalize' the local solutions. The presence of the negative cosmological constant renders the field equations regular at infinity. Work is ongoing on the stability of these solutions under time-dependent, $S$-wave perturbations, with indications that solutions with a sufficiently small electric field are stable.

\bigskip

The abstract of the talk by {\bf Gustavo Dotti} on \emph{Black hole non modal linear stability} can be found at the URL \url{gr20-amaldi10.edu.pl/index.php?id=49&abstrakt=284}.

\subsection{Extreme black holes}
 \label{ss16XI13.2}

In joint work  with J. Lucietti, K. Murata and N. Tanahashi~\cite{Lucietti:2012xr},
\textbf{Harvey Reall} addressed the question
\emph{What happens at the horizon of an extreme black hole?}
Reall explained how the already-mentioned Aretakis's instability~\cite{Aretakis4,Aretakis5}, of a massless scalar field at the horizon of an extreme Reissner-Nordstr{\o}m or extreme Kerr black hole, generalizes to any extreme black hole, and that a corresponding instability occurs for linearized electromagnetic and gravitational perturbations. Numerical simulations demonstrate that the instability persists when the gravitational back-reaction of the scalar field is included.

\bigskip

Next, \textbf{James Lucietti} considered the issue of \emph{Uniqueness of extreme horizons in Einstein-Yang-Mills theory}.
Any spacetime containing a degenerate Killing horizon, such as an extreme black hole, possesses a well defined  near-horizon geometry, as introduced by Reall~\cite{Reall:2002bh}. Such geometries have been the subject of much attention in modern studies of quantum gravity and high energy physics. In Einstein-Maxwell type theories in four and higher dimensions,  a variety of near-horizon uniqueness and symmetry enhancement theorems have been established~\cite{Kunduri:2013gce}. Less is known about extreme black holes and their near-horizon geometries when coupled to non-Abelian gauge fields.  It turns out that, for  Einstein-Yang-Mills theory with an arbitrary compact semi-simple gauge group (and allowing  a cosmological constant),  the near-horizon uniqueness and symmetry enhancement theorems persist~\cite{Li:2013gca}.
  Interestingly, in contrast to Einstein-Maxwell theory, a global analysis is required to establish the symmetry enhancement. This then allows one to prove that any (axisymmetric) near-horizon geometry with a compact horizon cross-section must be that of the Abelian embedded extreme Kerr-Newman black hole. The question of genericity of near-horizon rigidity remains.

\bigskip

\textbf{P\'eter Forg\'acs} talked about his work with
Peter Breitenlohner and Dieter Maison on \emph{Extremal black holes with regular interior in Einstein-Yang-Mills-Higgs theories}.
The authors studied gravitating magnetic monopoles in $\mathfrak{su}(2)$  Einstein-Yang-Mills-Higgs (EYMH) theory which
are static, spherically symmetric asymptotically flat solutions with regular origin~\cite{review}.
These solutions correspond to connecting trajectories between two hyperbolic fixed-points of the EYMH dynamical system depending on 2 parameters, the ``gravitational coupling strength'', $0\leq\alpha$, and the mass ratio $0\leq\beta$.
Analytical and numerical results show that when $\alpha\to\alpha_{\rm crit}$ the connecting trajectory (corresponding to a globally regular monopole) traverses a fixed point, decomposing thus into three parts; interior, exterior and the fixed-point itself; the latter corresponds to an AdS$_2\times$S$^2$ spacetime~\cite{BFM2}. The spacetime of the exterior part is that of a magnetically charged extremal Reissner-Nordstr{\o}m (RN)-type black hole.
Direct numerical construction of the critical solution
for values of $\beta$ up to $18$ provides convincing evidence for the existence of non-Abelian RN-type black holes whose interiors are regular.
Although the matter fields are not analytic at the horizon,
we find that they are at least $C^2$ for $\beta>12.1753$, and then
freely falling observers experience finite tidal forces when crossing the horizon.

\bigskip

The abstract of the talk entitled {\em
Analyticity of event horizons of extremal Kaluza-Klein black holes}, by {\bf
Masashi Kimura},  can be found at URL

\noindent
\url{gr20- amaldi10.edu.pl/index.php?id=49&abstrakt=751}.

\subsection{Miscellaneous}
 \label{ss16XI13.3}

Marginally Outer Trapped Surfaces (MOTSs) naturally belong to any discussion of black holes, as their occurrence
in spacetimes with conformal completions at infinity implies existence of black hole regions~\cite{CGS,galloway-nullsplitting}; see~\cite{AndMMS,AMS1,AMS2,CGP} and references therein for further results on MOTSs.
In his talk,
\textbf{Jos\'e-Luis Jaramillo} pointed-out some \emph{Physical aspects of MOTSs stability}.
 He reviewed the notion of outermost stability for marginally outer
trapped surfaces, and illustrated it in two applications admitting natural
physical interpretations. In the first one, MOTSs-stability provides
the key ingredient underlying a family of geometric inequalities setting
a lower bound for the MOTSs area in terms of angular momentum
and charges. This provides a notion of subextremality for black hole
horizons in dynamical situations with matter. The rigidity case is realized by slices of extremal
Kerr. When the ingoing expansion $\theta^{(k)}$ is negative
the surface gravity $\kappa^{(\ell)}$ must vanish~\cite{Jaramillo:2012zi}.
The second example provides a fluid analogy interpretation for the expression derived by
M. Reiris for the principal eigenvalue  of the MOTSs-stability operator
in axisymmetric stationary horizons, $\lambda_o = - \kappa^{(\ell)} \theta^{(k)}$,
in terms of a Young-Laplace law for ``soap bubbles''.
Denoting by $\gamma_{_{\mathrm{BH}}}$ the corresponding surface tension, by $\Delta p$
a formal ``pressure difference'' and by $H$ the MOTSs extrinsic curvature, one finds the correspondences~\cite{Jaramillo:2013rda}
\begin{eqnarray*}
 \frac{\lambda_o}{8\pi}\to \Delta p = p_{\mathrm{inn}}-p_{\mathrm{out}} \ ,
\frac{\kappa^{(\ell)}}{8\pi} \to \gamma_{_{\mathrm{BH}}} \ , \ -\theta^{(k)}\to H
\ \ \ \Longrightarrow \ \ \ \Delta p = \gamma_{_{\mathrm{BH}}} H \  .
\end{eqnarray*}

\bigskip

In her research,
\textbf{Mar\'ia Eugenia Gabach-Cl\'ement} has been interested in \emph{the force between axisymmetric black holes}.
The main issue is  existence of vacuum, stationary multiple-black hole configurations in axial symmetry.
Non-existence of such solutions with two components of the event horizon has been settled by Hennig and Neugebauer (see~\cite{HennigNeugebauer3,ChCo,CCH} and references therein), but the general case remains open.
One expects that there will be at least a conical singularity
on the portions of the symmetry axis between the different components of the horizon; this singularity can be interpreted as a   force between the black holes.  By examining the constraint equations, Gabach-Cl\'ement finds a lower bound on the force between the black holes,
in terms of the ADM mass, the angular momenta of the individual components and the distance. She also finds a quasi-local inequality for each black hole involving the area of the horizon,
the angular momentum and the charge, independently of the separation distance to the other black holes.

\bigskip

In joint work with
{Carlos Kozameh}  and Osvaldo Moreschi, \textbf{Emanuel Gallo}
reported on the results of \emph{A study of the field equations in the neighbourhood of a solitary black hole}. The authors developed a framework to analyze black holes in their late stage of dynamical evolution,  after all matter has fallen into the black hole, but taking into account both incoming and outgoing radiation. They propose a notion of a generalized surface gravity which is locally constant, consistently with the usual zeroth law of black hole mechanics.
The framework gives an alternative derivation of known results about linear perturbations around Schwarzschild black holes~\cite{KoMoPe}.
The goal is to analyze the Einstein field equations not only near a horizon $H$, but also in a region that contains null infinity, the horizon and timelike infinity, expecting to obtain more dynamical information on the late phase of gravitational collapse.

\bigskip

The abstract of the talk on {\em
Photon accumulation near a Schwarzschild black hole}
by
{\bf
Volker Perlick}
can be found at
\url{gr20-amaldi10.edu.pl/index.php?id=49&abstrakt=853}.

\section{Other}

In his talk on \emph{Dynamic and thermodynamic stability of perfect fluid stars},
\textbf{Stephen Green} explained how to extend the analysis by Hollands and Wald of dynamic
and thermodynamic stability of black holes and black branes
\cite{Hollands:2012sf} to the case of relativistic, perfect fluid
stars~\cite{Green:2013ica}.  The methods of~\cite{Hollands:2012sf} can
be carried over directly using a Lagrangian description of  perfect
fluids.  However, the perfect fluid case
differs in two major ways from that of black holes and black branes:
On the one hand, the absence of a black hole horizon eliminates many
of the difficulties encountered in~\cite{Hollands:2012sf}.  On the
other hand, the Lagrangian formulation of fluids requires the use of
potential fields, which introduces several new difficulties. Green, Schiffrin and Wald~\cite{Green:2013ica}
restrict their analysis to stars in {\em dynamic equilibrium}: stationary,
axisymmetric solutions of the Einstein-fluid equations with circular
flow.  They define notions of {\em thermodynamic equilibrium}, as well
as {\em dynamic} and {\em thermodynamic stability}, and prove
various results relating these notions.  In particular,   a {\em
  canonical energy} $\mathcal E$ is used to develop a linear
stability criterion for a certain class of axisymmetric perturbations.
For Lagrangian perturbations of stars in thermodynamic equilibrium, it is
shown that a ``canonical energy in the rotating frame'' $\mathcal E_r$ satisfies
$$
 \mathcal{E}_r=\delta^2M - \Omega \delta^2J
 \;,
$$
%.
and that for
axisymmetric perturbations $\mathcal{E}=\mathcal{E}_r$.

\bigskip

One expects that the analysis by
\textbf{Nils Andersson} of \emph{A covariant action principle for dissipative fluid dynamics}
could provide a starting point of further applications of the study just described.
In his talk on that topic,  Andersson presented a new variational framework for dissipative general relativistic fluid dynamics. The model extends the convective variational principle for multi-fluid systems to account for a range of dissipation channels. The key ingredients of the construction are i) the use of a lower dimensional matter space for each fluid component, and ii) an extended functional dependence for the associated volume forms.  The new formalism impacts on both applications and foundational issues.

\bigskip

In his talk,
\textbf{Jacek Tafel} presented the results of his search,  with Michal J\'o\'zwikowski~\cite{TJ}, of a \emph{Generalization of initial data for the Kerr metric}. The authors consider axisymmetric and asymptotically flat initial data sets  $(g,K)$ with  $K^i_{\ i}=0$, where the Killing vector $\partial_{\varphi}$ has no twist.
 The general solution of the momentum constraint for such metrics is obtained.  One can then use a conformal rescaling to solve  the Hamiltonian constraint. Variants of this approach are considered where marginally trapped surfaces are incorporated into the scheme (see~\cite{Maxwell:ah,Dilts:2013hoa,MeierHolstTsogtgerel} for further results on this last issue).

\bigskip

The title of the talk of \textbf{Alfonso Garc\'{\i}a-Parrado G\'omez-Lobo} was \emph{On the characterization of non-degenerate foliations of pseudo-Riemannian manifolds with conformally flat leaves}. The speaker presented a characterisation of foliations of pseudo-Riemannian
manifolds with conformally flat leaves in terms of vanishing of certain tensors. These tensors are defined using a new affine connection called the {\em bi-conformal connection}~\cite{BICONF-1} whose geometric
relevance in the study of conformally separable pseudo-Riemannian manifolds was noticed in~\cite{BICONF-1}.
The bi-conformal connection is defined in terms of the projection morphisms
determined by the leaves of the foliation and the Levi-Civita connection of the pseudo-Riemannian metric. The case where the leaves have co-dimension three plays a distinguished role. A  restrictive condition is that
the leaves must be non-degenerate hypersurfaces~\cite{BICONF-3}.

\bigskip

The reader interested in an {\em
Almost Birkhoff Theorem in General Relativity}
will find the abstract of the talk by
{\bf
Rituparno Goswami}
on
 \url{gr20-amaldi10.edu.pl/index.php?id=49&abstrakt=487}.

\bibliographystyle{amsplain}

\bibliography{GR20_A2sessionNEW}

\end{document}